\documentclass[letterpaper,twocolumn,10pt]{article}
\usepackage{usenix2019_v3}

\usepackage[justification=centering]{caption}
\usepackage{tikz}
\usepackage{amsmath}

\usepackage{filecontents}
\usepackage{times}
\usepackage{fancyhdr,graphicx,amsmath,amssymb}
\include{pythonlisting}
\usepackage{float}
\usepackage{graphicx}
\usepackage{url,hyperref}
\usepackage[mathscr]{euscript}
\usepackage[linesnumbered,vlined,ruled,boxed,commentsnumbered]{algorithm2e}
\usepackage{algpseudocode}
\usepackage{enumerate}
\usepackage{paralist}
\usepackage{forest}
\usepackage{adjustbox}
\usepackage{xr}
\usepackage{mathtools}
\usepackage{amssymb}
\usepackage{amsthm}
\usepackage{siunitx}
\usepackage{booktabs}
\usepackage{svg}

\usetikzlibrary{arrows}
\tikzset{
	vertex/.style={circle,draw,minimum size=1.5em},
	edge/.style={->,> = latex'}
}
\usepackage{subcaption}
\usepackage{bm}
\usepackage{tabularx}
\begin{document}

\date{}

\title{\Large \bf DeePref: \underline{Dee}p Reinforcement Learning For Video \underline{Pref}etching In Content Delivery Networks}
\author{
{\rm Nawras Alkassab}\\
University of South Carolina
\and
{\rm Chin-Tser Huang}\\
University of South Carolina
\and
{\rm Tania Lorido Botran}\\
Roblox
}

\maketitle


\begin{abstract}
Content Delivery Networks carry the majority of Internet traffic, and the increasing demand for video content as a major IP traffic across the Internet highlights the importance of caching and prefetching optimization algorithms. Prefetching aims to make data available in the cache before the requester places its request to reduce access time and improve the Quality of Experience on the user side. Prefetching is well investigated in operating systems, compiler instructions, in-memory cache, local storage systems, high-speed networks, and cloud systems. Traditional prefetching techniques are well adapted to a particular access pattern, but fail to adapt to sudden variations or randomization in workloads. This paper explores the use of reinforcement learning to tackle the changes in user access patterns and automatically adapt over time. To this end, we propose, \textsc{DeePref}, a Deep Reinforcement Learning agent for online video content prefetching in Content Delivery Networks. \textsc{DeePref} is a prefetcher implemented on edge networks and is agnostic to hardware design, operating systems, and applications. Our results show that DeePref DRQN, using a real-world dataset, achieves a 17\% increase in prefetching accuracy and a 28\% increase in prefetching coverage on average compared to baseline approaches that use video content popularity as a building block to statically or dynamically make prefetching decisions. We also study the possibility of transfer learning of statistical models from one edge network into another, where unseen user requests from unknown distribution are observed. In terms of transfer learning, the increase in prefetching accuracy and prefetching coverage are [$30\%$, $10\%$], respectively. Our source code will be available on Github \footnote{https://www.github.com}.

\end{abstract}

\section{Introduction and Related Work}
\label{sec:Introduction and Related Work}


Prefetching is a well-studied speculation technique that aims to make data available in the cache before the requester places their request. A prefetcher masks/reduces I/O latency between the data provider and consumer by predicting future demands based on historical access patterns. Prefetching differs from caching \cite{shafiq2014revisiting, pallis2006insight, pathan2007taxonomy, rimal2009taxonomy} in the sense that data needs to be available in the cache before the initial request by predicting future accesses and fetching them into memory with the hope that they will be requested in the near future. In caching, the item needs to be requested at least once in order for it to be cached. Substantial work has been done studying the interaction between prefetching and caching \cite{cao1995study, patterson1995informed, kimbrel2000near}. For example, Kimbrel and Karlin \cite{kimbrel2000near} had made the first attempt towards understanding prefetching and caching pages from parallel disks. The authors provided an offline theoretical model that captures the characteristics of the system for prefetching with multiple disks with a limited look-ahead into the request stream. Such an oracle is often not provided, but it is helpful for setting up a baseline and understanding the integration between caching and prefetching. \\

\begin{figure}[h]
\centering
\includegraphics[width=0.5\textwidth]{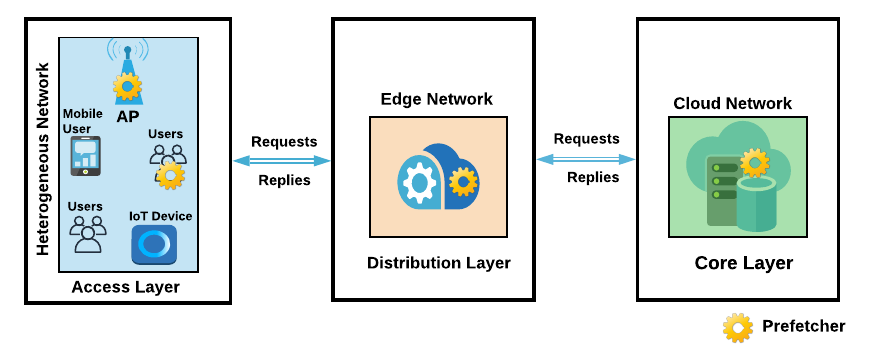}
\caption{Content Delivery Networks Architecture with possible prefetching implemented in different nodes across different layers. }
\label{fig:CDNs}
\end{figure}

Many prefetching techniques have been proposed over the last two decades to exploit hardware design/features \cite{vander1999compiler, guo2007compiler, zucker2000hardware, vander1997caches, mittal2016survey, bera2021pythia}, compiler-injected instructions \cite{chen1995effective, lattner2005automatic, rabbah2004compiler, peled2015semantic, kolli2013rdip, ferdman2011proactive}, or memory access patterns in operating systems \cite{voelker1998implementing, al2020effectively}. They are often limited to specific hardware design, access pattern, or application-specific features/hints, and they exploit the design of a lower level memory stack. For example, LEAP \cite{al2020effectively} is a kernel prefetcher designed specifically for applications using remote memory. Such design of prefetching algorithms limits its applications to various workloads which exhibit a mixture of both sequential and random streams. In this work, we present a prefetching algorithm that is agnostic to the hardware design, operating system, and applications/workloads. Our approach takes only the request ID to decide when and what to prefetch. \\

Furthermore, prefetching applications have become a prevalent feature in numerous applications, such as local memory management, wherein the item is prefetched from a Hard Disk Drive (HDD) or Solid State Drive (SSD) to the local cache. \textsc{SARC} \cite{gill2005sarc}, for example, is a prefetching algorithm that dynamically partitions the cache space into sequential blocks representing the prefetched items and random blocks representing the on-demand pages. The \textsc{SARC} algorithm considers the cache pollution by equalizing the marginal utility of both prefetched data and demand-paged data.  This equalization of cache misses between random blocks and sequential blocks is not quite fair, since sequential blocks require less time to fetch compared to random blocks due to their spatial locality exposed in HDDs and SSDs. Therefore, a more weight should be given to penalizing fetching random blocks. \textsc{DULO} \cite{jiang2005dulo} had addressed this problem by giving random blocks more weight for being kept in cache to compensate for their prefetching cost. In \textsc{DULO}, the characteristics of the hard disk are exploited so that sequential access is more efficient than random access. Further, \textsc{AMP} \cite{gill2007amp} reduces cache pollution by determining the prefetching trigger point (i.e., when to prefetch) and the prefetching degree (i.e., amount of data to prefetch). \textsc{QuickMine} \cite{soundararajan2008context} is a context-aware online prefetching technique that relies on hints from applications to discover block correlations in storage systems. \textsc{QuickMine} is a history-based prefetching technique which is typically expensive compared to sequential prefetching algorithms. To overcome that, \textsc{QuickMine} tags each application I/O block request with a context identifier corresponding to a higher level application context (e.g., a web interaction, database transaction). The tag enables the request sequence to be split before mining, thus masking computation overhead is manageable. The key novelty of \textsc{QuickMine} lies in detecting and leveraging block correlations within application contexts. However, \textsc{QuickMine} leverages the application contexts to make prefetching predictions. Thus, \textsc{QuickMine} requires insights/hints from applications which may not be available or feasible to the prefetching head in some environment. Rendering \textsc{QuickMine} unavailable for legacy systems and makes it hard to deploy. VanDeBogart \textit{et al.} \cite{vandebogart2009reducing} proposed, \textsc{libprefetch}, an application-directed prefetching scheme that aims at minimizing the I/O fetch time of non-sequential single-disk reads by speculating the access patterns deduced from the application source code itself.  Chang \textit{et al.} \cite{chang1999automatic} focused on parallel disk I/O systems by providing more I/O bandwidth. For SSD prefetching, the authors of \textsc{FAST} \cite{joo2011fast} proposed a prefetching scheme on SSD drives to accelerate the application launch time. Their approach is inherited from the fact that the I/O request sequence does change over repeated start-up of the application during cold start, by overlapping the SSD access (I/O) time with the computational (CPU) time by running the application prefetcher concurrently with the application itself. \\

 \begin{figure}[h]
     \includegraphics[width=0.5\textwidth]{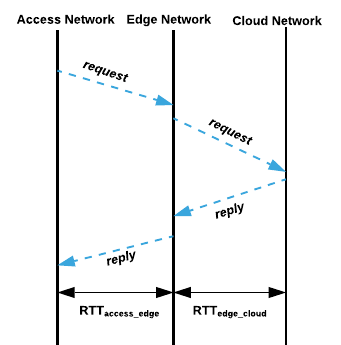}
 	\caption{An illustrative flow chart of the communication between access network, edge network, and cloud network upon a cache miss in CDNs. \textbf{Note: } RTT between edge and cloud networks is much larger than RTT between access and edge networks.} 
     \label{fig:flow_chart}
 \end{figure}
Furthermore, hardware data prefetching is a well-known technique for hiding and tolerating off-chip memory latency for single core and multicore systems \cite{albericio2012abs, panda2015caffeine, panda2016spac, sun2019combining, bera2021pythia, zhu2020ctdgm}. Recently, Bera el al. \cite{bera2021pythia} proposed a hardware prefetching technique, \textsc{Pythia}, which learns to prefetch using multiple different types of program context and system-level feedback information inherent to its design by formulating the problem as Markov Decision Process (MDP) and treating the prefetcher as a reinforcement learning agent. We take a similar approach in this work by formulating the prefetching problem in CDNs as a reinforcement learning problem, which we will discuss in Sec.\ref{sec:System Architecture}. For a recent survey of hardware prefetching, please visit this paper \cite{mittal2016survey}. For web applications and cloud tenants prefetching, block storage systems such as cloud tenants are disk-bound where the bottleneck is accessing blocks from the disk itself (i.e., HDD, SSD), recent caching and prefetching techniques for cloud applications have been proposed \cite{waldspurger2015efficient, yang2017mithril, jiang2013prefetching, yang2016tombolo, chiang2015adaptive} to mitigate access latency. For example, \textsc{MITHRIL} tracks temporal associations between blocks whose access patterns are moderately frequent, leaving the more frequent requests to be handled by the underlying caching mechanism. Unlike \textsc{MITHRIL}, our approach does not rely on the underlying caching system by eliminating the caching head and leaving the prefetching head to accomplish prefetching tasks only. \\

For web security, a full TLS handshake between HTTPS client and web server is considered a bottleneck, since the client needs to fetch and validate the server's certificate before a secure connection between client and server is established. A typical TLS handshake consists of two round trips to complete. Caching the server's certificate on the web browser reduces the handshake to one round trip \cite{shacham2002fast, langley2016transport} or even to zero round trips \cite{langley2010transport}, however, the TLS certificate's validation status may not be cached, and the latency imposed by the validation process will still impact the QoE on the client side. Therefore, Stark \textit{et al.} \cite{stark2012case} proposed a prefetching and pre-validating scheme for TLS server certificate into the client's web browser, allowing the client to perform the TLS handshake and validate the server's certificate with zero round trips. Further, one of the methods proposed to perform certificate prefetching is to prefetch the server certificate from CDNs to avoid any overload on the server side \cite{stark2012case}. \\

Another interesting line of prefetching is web content prefetching \cite{summers2014automated}. Summers \textit{et al.} \cite{summers2014automated} proposed an adaptive prefetching scheme for HTTP streaming video servers that automatically adjusts prefetch size to gain higher disk throughput on the server side. To do so, their algorithm uses system measurements such as available memory, disk usage, and video bit rate from the operating system to inform the application to periodically adjust the prefetch size. The authors treat the entire operating system as a gray box and provide application-level prefetching implementation without modifying the kernel. Further, extensive experimentation on HTTP Streaming video workloads had been conducted previously \cite{summers2016characterizing, summers2012methodologies} to better design flexible prefetching algorithms (i.e., aggressive vs. non-aggressive). Measurement studies \cite{breslau1999web, gill2007youtube} show that video content has a long tail popularity distribution similar to Zipf distribution. Meaning that only few of the hottest video files will benefit from in-memory caching. However, studies show that 70-80\% of videos are watched only once \cite{gill2007youtube, zink2009characteristics}. Therefore, we exclude, in this work, caching from the memory management system on edge devices to allow the edge node to accomplish prefetching tasks only as discussed in Sec.\ref{sec:System Architecture}. Moreover, assuming that the workload follows Zipf-like distribution throughout the entire experiment may be invalid, since the popularity distribution of video content may change very rapidly in CDNs. Thus, we only provide visibility on the popularity distribution to our prefetching agent as part of its initial state. \\

For high-speed networks with low latency, Voelker et al. \cite{voelker1998implementing, bartels1999potentials} introduced a prefetching and caching algorithm that utilizes the free memory in idle workstations in Local Area Networks (LAN) environments whereas the cost of prefetching over a high-speed network is less than the cost of prefetching locally on the hard-disk. This assumption is invalid considering CDNs because the propagation delay is considerably large compared to LAN environments. Another strong assumption the authors had made is that their central algorithm assumes a complete prior knowledge of the reference streams of the applications running on all participating nodes, including the pages to be referenced, the relative order in which they are referenced, and the inter-reference times. These assumptions cannot hold true considering nowadays applications and workloads that express more stochasticity than sequentiality such as cloud-based applications and video streaming applications in CDNs. \\

Recently, Multi-access Edge Computing (MEC) enabled caching and prefetching techniques to be implemented at edge networks, enhancing QoE at the user-side while benefiting cloud server at the core layer in CDNs. Video streaming applications can utilize MEC environments to prefetch video segments to end users at edge networks, allowing multi-users that are, possibly, close geographically to benefit from the prefetched segments. \textsc{SPACE} \cite{aguilar2023space} is a prefetching and caching framework that utilizes MEC to prefetch video segments to edge networks in cellular networks (i.e., 5G). \textsc{SPACE} models video segment prefetching as a Markov finite state machine which can be updated online and it uses the buffer size, link bit rate, previous QoE, previous link bit rate. The authors of \textsc{SPACE} also proposed a cache eviction algorithm to evict the prefetched segments, named Least Popular Used (LPU). Behravesh \textit{et al.} \cite{behravesh2022machine} proposed a segment prefetching algorithm based on bit rate prediction of client segment requests using a near-optimal machine learning approach. \textsc{COOPEC} is a video segment prefetcher that utilizes client requests and past viewing history to minimize the bit rate oscillations caused by the web browser when cache hits are inconsistent. \textsc{MVP} \cite{ge2017toward} is a video segment prefetching technique implemented at edge networks for 4K Videos On Demand (VoD) applications. \textsc{MVP} achieves seamless playback in realistic LTE-A network infrastructure. \\

To this end, we hypothesize that prefetching an entire video content with high quality has its advantages over segment prefetching in CDNs where the bottleneck is the upstream network between edge networks and cloud networks. To name a few, 1) Prefetching video content as a bulk saves resources in the upstream network, however, segment prefetching is still subject to fluctuations in the upstream network since a continuous Transmission Control Protocol (TCP) session needs to be maintained throughout watching the video content. 2) Round Trip Time (RTT) of edge-cloud network is much larger than access-edge network due to the propagation delay, therefore, prefetching an entire video content eliminates any propagation delay of future requests. 3) QoE is guaranteed in bulk prefetching, and it is only subject to Adaptive Bit Rate Algorithms (ABR) of the end user network (i.e., access networks). To illustrate, Fig.\ref{fig:flow_chart} shows a flow chart of user requests generated in access networks and forwarded to edge networks. Upon a cache miss, the request will be forwarded to the cloud network in CDNs. We design a prefetcher that resides at edge networks and makes prefetching decisions synchronous to user requests online. The aim is to prefetch an entire video content with high quality, with the hope that future requests from the same or different user will occur soon. Our approach is minimally invasive and is agnostic to hardware design, operating system, and application. \\

\begin{figure}[h]
    \includegraphics[width=0.5\textwidth]{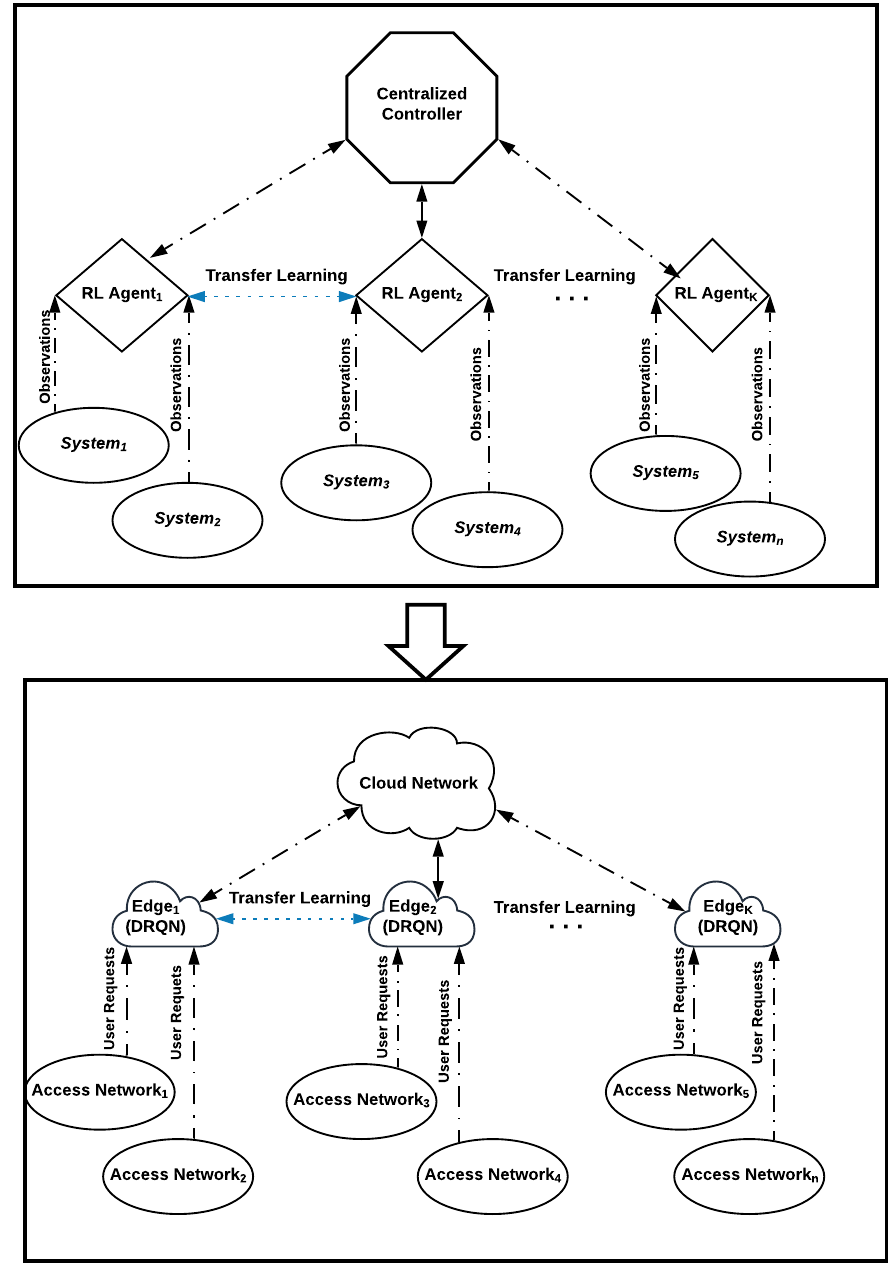}
	\caption{Learning Local prefetching policies on edge networks.} 
    \label{fig:local_policy}
\end{figure}

In this work, we utilize both \textsc{DQN} and \textsc{DRQN} designs as building blocks to tackle the prefetching problem. We propose a prefetcher which is implemented at edge networks where possible transfer learning can occur between edge networks, as shown in Fig,\ref{fig:local_policy}. We discuss the results for transfer learning in Sec.\ref{sec:results}, but first we express the need to use Deep Reinforcement Learning (DRL) due to the large state and action spaces exposed in video content prefetching in CDNs. \\

Our results show that \textsc{DeePref DRQN}, using a real-world dataset, achieves 17\% increase in prefetching accuracy and 28\% increase in prefetching coverage on average compared to baseline approaches that use video content popularity as a building block to statically or dynamically make prefetching decisions. In terms of transfer learning, the increase of prefetching accuracy and prefetching coverage are [$30\%$, $10\%$], respectively. We discuss the experimental design and results in Sec. \ref{sec:Experiment Design}.
We show that \textsc{DeePref DRQN} outperforms baseline approaches with an increase of at least [$17\%$, $28\%$] in terms of prefetching accuracy and coverage, respectively. \\


\section{Motivation And Contribution}
\label{sec:Motivation And Contribution}

In recent years, we have witnessed an exponential growth in Internet traffic. Mainly, this is due to the explosion of Big Data \cite{mcafee2012big}, the increased usage of social media \cite{kaplan2010users}, the emergence of cloud computing \cite{mell2011nist} and edge computing \cite{shi2016edge}. This rapid growth has accelerated the implementation and design of CDNs. Cisco \cite{cisco2018cisco} had predicted that around 64\% of the entire Internet traffic will cross CDNs in 2020, up from 45\% in 2015. Video streaming service providers such as Netflix \footnote{https://www.netflix.com} and Hulu \footnote{https://www.hulu.com} rely heavily on CDNs to distribute their video content to end users \cite{adhikari2014measurement}. For example, Akamai Technologies \footnote{https://www.akamai.com} is the market leader (80\% of the overall CDN market) in providing content delivery services where it owns more than 12,000 servers over 1,000 networks in 62 countries \cite{pallis2006insight}. Furthermore, There exists an increased trend in the popularity of video content as a major internet traffic. According to Cisco \cite{cisco2018cisco}, global Internet video traffic (business and consumer, combined) will grow 4-fold from 2015 to 2020, a compound annual growth rate of 31\%. Internet video traffic will be 79\% of all Internet traffic in 2020, up from 63\% in 2015. In addition, Internet video traffic will reach 127.8 Exabytes per month in 2020, up from 33.7 Exabytes per month in 2015, and Ultra HD and HD video content will be 15.7\% of Internet video traffic in 2020, up from 2.3\% in 2015. This recent trend of video popularity highlights the importance of video content prefetching from cloud networks to fog networks in CDNs. Further, prefetching can be implemented in various nodes from end users to cloud servers in CDNs, as illustrated in Fig.\ref{fig:CDNs}. Recently, Hu \textit{et al.} \cite{hu2017musa} proposed a Reinforcement Learning (RL) approach to prefetch successive episodes of video series to access points that are shipped with large storage capacity and close to end users, the assumption is that the end user is more likely to watch more than one episode without skipping. \textsc{IPAC} \cite{liang2015integrated} integrates prefetching with caching to allow the video player at the end user side to generate prefetch requests for successive video segments with the same bit-rates as the current request. \textsc{IPAC} requires a modification of the underlying cache eviction algorithm, and it does not consider multi-user prefetching (inter-correlation among user requests). Further, Alkassab \textit{et al.} \cite{alkassab2017benefits} addressed the advantages and the disadvantages of video content prefetching across different layers (core, distribution, and access) in CDNs and discussed the utilization of out-of-band bandwidth for prefetching in fog networks. Wu \textit{et al.} \cite{wu2021optimization} proposed a prefetching scheme based on Bayesian Networks and Markov chains to prefetch video files and designed a cache eviction algorithm which utilizes the content heat and re-access probability such that the cached content with the least re-access weight is evicted. On the other hand, Sham \textit{et al.} \cite{sham2021intelligent} proposed an admission control system to place prefetching requests based on CPU, memory and other resource parameters in hybrid edge-cloud networks. Such design of prefetching algorithms require system-level and application-level information that may not be available at the prefetcher side, limiting the applicability/implementation of such algorithms.\\

As discussed previously, prefetching in CDNs has many advantages such as preventing bandwidth under-utilization, optimize server load, and reducing access latency. Consequently, improving the overall system response time and the Quality of Service (QoS) experienced by end users. On the other hand, prefetching increases system resource demands in order to improve response time which increases the incurred bandwidth consumption and storage resources \cite{venkataramani2002potential}. Thus, prefetching in CDNs exposes two difficult challenges that can be addressed by the following questions:
\begin{itemize}
    \item What items should be prefetched, considering the large volume of video content residing on video streaming platforms (e.g., Netflix) and millions of end users connected to thousands of edge devices?; a question that is similar to what items to suggest to end users by recommendation systems.
    \item And, when a prefetching action should occur?; prefetching items very frequently indicates more eviction actions to be taken by the edge network which increases cache pollution, network traffic, and server load at the cloud network, hence, reducing the overall system performance. On the other hand, prefetching items on rare occasions reduces the prefetching coverage and accuracy, since content prefetching is a time-sensitive subject. 
\end{itemize}

We propose, \textsc{DeePref}, a deep reinforcement learning agent for prefetching video content from cloud networks to edge networks in CDNs. \textsc{DeePref} is agnostic to hardware design, operating system, and application where it takes only the request ID of the user request as input to make future prefetching decisions online. \\

In a separate line of research, Reinforcement Learning (RL) \cite{sutton2018reinforcement} has recently witnessed a tremendous success in games \cite{silver2016mastering, tesauro1995temporal} as well as in robotics \cite{kober2013reinforcement, levine2016end}. RL focuses on building agents that interact with an environment to maximize a cumulative future reward or a long-term reward. The environment or the problem whom the agent is trying to solve needs to constitute the Markovian property in order to be modeled as a Markov Decision Process (MDP) \cite{puterman2014markov}, as in Def. (\ref{def:MDP}). Unfortunately, many real-world problems do not directly exhibit the Markovian property and/or the agent can only \textit{partially observe} the underlying states of the environment, prohibiting it from discovering the \textit{true} states of the environment. Thus, a generalization of MDPs is called Partially Observable Markov Decision Processes (POMDPs) allow us to learn decision policies if it's assumed that the underlying environment is Markovian (e.g., MDP). In this direction, we frame the prefetching problem as an MDP, as in Def. (\ref{def:MDP}) and investigate different design and implementation approaches to learn prefetching policies in CDNs where possible transfer learning can occur from one edge network into another. \\

\theoremstyle{definition}
\newtheorem{definition}{Definition}[section]

\begin{definition}{Markov Decision Process or MDP}
\label{def:MDP} is defined as a tuple $M := (S, A, P, R, \rho_{0}, \gamma$). Where $S$ is the state space, $A$ is the action space, and $P$ is the transition function from state $s$ to state $s^{'}$ after taking an action $a$, that is ($s \xrightarrow[]{a} s^{'}$). The reward function $R : S \times A \times S \rightarrow \mathcal{R}$ is a random variable representing the reward after transitioning from one state to another and observing a reward $R_t = R(S_{t}, A_{t}, S_{t+1})$, $\rho_{0}$ is the initial state distribution, and $\gamma \in [0, 1]$ is the discount factor that trades off the instantaneous and future rewards. 
\end{definition}





\section{System Architecture}
\label{sec:System Architecture}
\begin{figure}[h]
\centering
\includegraphics[width=0.5\textwidth]{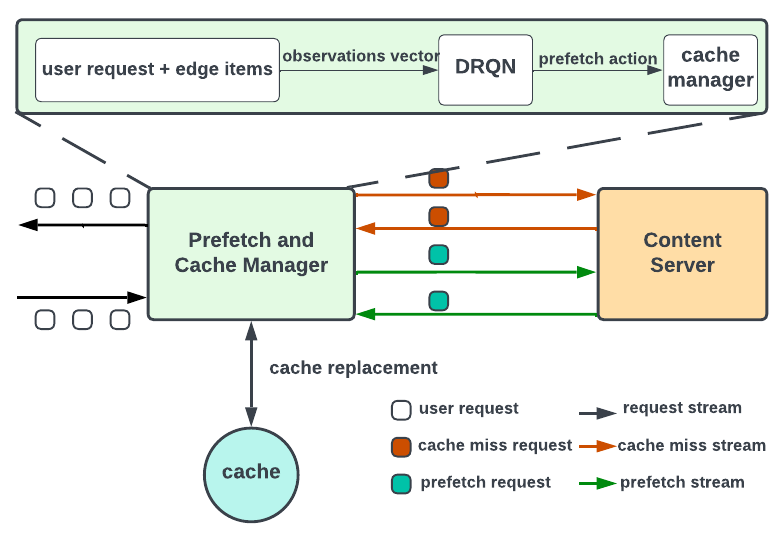}
\caption{DeePref System Architecture.}
\label{fig:system_architecture}
\end{figure}


We design, DeePref, such that a prefetching decision is made at edge networks every time a user request is received from access networks. Consequently, the prefetching requests are synchronous to user requests as shown in DeePref architecture, see Fig.\ref{fig:system_architecture}. We also eliminate the cache manager from the cache management system and allow the prefetch manager to accomplish prefetching tasks only, unlike other approaches such as IPAC \cite{liang2015integrated}. We use LRU for cache eviction when comparing DeePref to baselines approaches unless otherwise mentioned. \\

Further, the set of video content residing at content servers in cloud networks can be thought of as a set of actions the prefetcher can take at edge networks. Therefore, we add a new action which is the "do-not-prefetch” action in order to tolerate the aggressiveness of our approach. In fact, our results show that DeePref is more moderate compared to baselines, as discussed in Sec.\ref{sec:results}. Moreover, we assume that the cloud network periodically updates the edge network with newly added video content to accommodate for prefetching new/unseen items. We also assume that the connection between the cloud network and edge network is the bottleneck considering the limited bandwidth in the upstream network. In other words, we assume that the Round Trip Time (RTT) of edge-cloud network is much larger than RTT of access-edge network. \\

\textsc{DeePref} prefetches video content at the same bit-rate since there is high probability (> 0.83) that a video player residing at the end user side will send the next request with the same bit-rate as the current request \cite{liang2015integrated} because the video play avoids frequent bit-rate changes which degrades the QoE at the end user side \cite{cranley2006user, jiang2012improving}. \\

\section{Learning Prefetching Policies On Edge}
\label{sec:Learning Prefetching Policies On Edge}
In its most abstract setting, RL aims to find an optimal policy for an agent sequentially interacting with an environment over a sequence of time-steps, $\mathit{t = 1, 2, 3, ...,}$. The agent executes actions and receives observations and rewards, and the aim is to maximize the future cumulative reward \cite{russel2013artificial}. Value-based approaches such as Q-learning \cite{watkins1992q} and policy-based ones (policy-gradient) such as REINFORCE \cite{williams1992simple} constitute classification approaches to solve RL problems \cite{sutton2018reinforcement}. Value-based approaches present many advantages such as seamless off-policy learning and better sampling efficiency compared to policy-gradient methods, however, they are prone to learning instability when mixed with function approximation \cite{sutton1999policy}. Although policy convergence in value-based methods is not well-studied, the emergence of deep learning \cite{lecun2015deep} has empowered the recent success of many value-based methods such as Deep Q-networks (DQN) \cite{mnih2013playing} and DRQN \cite{mnih2013playing}. \\

In this work, we follow the norm of decoupling the functionality of the underlying cache management system from the prefetching head, and we eliminate the cache management system which allows only the prefetching head to make prefetching decisions, as shown in Fig.\ref{fig:system_architecture}. Further, we use Least Recently Used (LRU) as a unified eviction algorithm when comparing prefetching policies, unless otherwise mentioned. We also assume no partial prefetching when making prefetching decisions since prefetching successive video content by means of partial prefetching may leave many prefetched blocks inaccessible, which increases the prefetching pollution. Our work considers prefetching one video content as a whole and the aim is to maximize both prefetching accuracy and prefetching coverage at the edge node by prefetching video content that have the highest likelihood to be accessed by end users since each video content may be accessed at least once \cite{gill2007youtube, zink2009characteristics}. \\

\subsection{Large State And Action Space}
\label{sec:Large State Space}
If we decide to prefetch one video content naively from the cloud network to an edge network, the problem of prefetching can be thought of as selecting one of the available video content among all content residing on the cloud network. Thus, we can model the prefetching problem as a Markov Decision Process (MDP) \cite{puterman2014markov} as in Definition, \ref{def:MDP}, which is a tuple $(S, A, P, R, \gamma)$. Where $S$ is the state space, $A$ is the action space, $P$ is the transition function, $R$ is the reward function, and $\gamma$ is the discount factor $\gamma \in [0,1]$. Consequently, we can construct the underlying MDP as follows.
\begin{itemize}
  \item \textbf{\emph{State Space: }} 
  Let $S= \{s_1, s_2, ..., s_t, ..., s_T\}$ denote the state space. Such that the current state at time step $t$ can be represented as: $s_t =\{m_1, m_2, \dots, m_i, m_M \} \in \mathbb{R}^{C_e}$ where $m_i \in M$, $M$ is the total number of video content residing at the cloud network, and $C_e$ is the storage capacity at the edge node $e$.
  
  Intuitively, the state dimensions/space can be expressed through the following equation:
  \begin{equation}
  \label{eq: equation_1}
  \mid S \mid = \binom{M}{C_e},
  \end{equation}

  \item \textbf{\emph{Action Space: }}
  Let $A = \{a_1, a_2, ..., a_t, ..., a_T\}$ denote the action space. Where $a_t$ is the prefetching action to be taken by the cloud device at time step $t$. The cloud device can make the following possible prefetching decisions:
  \begin{enumerate}
      \item \textbf{\emph{Prefetch one video content: }}
      The edge node will choose to prefetch one video content among the available $M$ video content located at the cloud.

      \item \textbf{\emph{No prefetching action required: }}
      The edge node $E$ can decide not to prefetch at all. Consequently, there will be no eviction of items taken at the edge node $E$.
  \end{enumerate}

  Consequently, the action space can be expressed as follows.
  \begin{equation}
  \label{eq:single_RL_action_space}
      \mid A \mid = \binom{M}{1} + 1,
  \end{equation}
  
  \item \textbf{\emph{Transition Function: }}
  At time step $t+1$, the system state transitions to another state after taking an action where the transition probability can be given by the following equation:
  \begin{equation}
      P(s_{t+1} \mid s_t, a_t)
                              = P(m\textsuperscript{e}_{t+1} \mid  m_{t}\textsuperscript{e}, a_t),
  \end{equation}

\end{itemize}

Thus, the system's state/action space can be expressed as follows:
  \begin{equation}
      \mid S \mid \cdot \mid A \mid = (\binom{M}{1}+1) \cdot \binom{M}{C_e},
  \end{equation}

Where $M$ is the total number of video content residing at the cloud network, and $C_{e}$ is the storage capacity at edge network, such that $M \gg C_{e}$. The state space (e.g., how many possible combinations of item that can reside at edge network is given by $\mid S \mid = \binom{M}{C_e}$, and the action space is $\mid A \mid = \binom{M}{1} + 1$ (e.g., selecting one video content to prefetch or decide not to prefetch at all). \\

Naturally, the prefetching problem suffers from large state space and action space and deriving an optimal prefetching policy is tricky since the action space is non-stationary and items are continuously added to the cloud network (e.g., continuous action space). Moreover, an optimal prefetching policy differs from an optimal caching policy such that in prefetching the item selection is not constrained to the set of items that had been previously seen (e.g., cached) by edge networks, rather, an optimal prefetching policy needs to \textit{consider/predict} any item to be requested despite how many times it had been requested before. Unlike caching, where an optimal policy can be derived by selecting one of the items that had been previously cached before, which reduces the action space and problem dimensionality. \\ 

\subsection{Reward Design}
\label{sec:Reward Design}
 As mentioned previously, the heart of reinforcement learning is to learn a reward function to maximize the agent's future cumulative reward. This learning mechanism depends heavily on the reward design. In \textsc{DeePref}, we simplify the reward design as much as possible to entail all aspects of prefetching in numerical values ranging between $[-2, 2]$ based on normalized latency, $[0, 1]$ as described below:
 \begin{itemize}
    \item The agent receives a reward of $2$ if there is a cache hit and the agent decided not to prefetch. This indicates that the item already resides at edge and had been prefetched before.
    
    \item The agent receives a reward of $2 - l_i$ if there is a cache hit and the agent decided to prefetch an item $i$ with latency $l_i$. 
    
    \item The agent receives a reward of $-1$ if there is a cache miss and the agent decided not to prefetch. 

    \item The agent receives a reward of $-1 - l_i$ if there is a cache miss and the agent decided to prefetch an item $i$ with latency $l_i$.

\end{itemize}

The goal of a single agent \textit{Reinforcement Learning} (RL) is to solve the underlying MDP by finding a policy $\pi : S \rightarrow \Delta(A)$, a mapping from the state space $S$ to a distribution over the action space $A$, so that $a_{t} \sim \pi(\cdot|s_{t})$ and maximizing the return $G$. The \textit{return} is the discounted accumulated reward, that is, $G_{t} = \sum_{t=0}^{\infty} \gamma^{t} R(s_{t},a_{t},s_{t+1})$. We can define a \textit{value function} $V(s)$ Eq. (\ref{eq:value_function}) or \textit{action-value function} $Q(s,a)$ Eq. (\ref{eq:action_value_function}) under policy $\pi$ as follows:

\begin{equation}
    \label{eq:value_function}
    V(s) := E [ G_t \mid S_t = s],
\end{equation}
\begin{equation}
    \label{eq:action_value_function}
    Q(s,a) := E [ G_t \mid S_t = s, A_t = a ],
\end{equation}

In Reinforcement Learning (RL), we usually use methods of \textit{Dynamic Programming (DP)} \cite{puterman2014markov}, to maximize the return $G_t$ by calculating a \textit{value function} $V(s)$ or  the \textit{action-value function} $Q(s,a)$ when using policy $\pi$. The computations carried in this procedure to approximate an optimal policy $\pi^{*}$ is called \textit{policy improvement} and \textit{policy evaluation} \cite{sutton1998introduction} where we typically use \textit{value-based methods} or \textit{policy-based methods} such as \textit{Temporal Difference} (TD) learning algorithms \cite{tesauro1995temporal, tsitsiklis1997analysis, sutton1998introduction} which have convergence guarantees under some conditions. For example, Q-learning \cite{watkins1992q} is an \textit{off-policy} TD learning algorithm that may converge to an optimal policy $\pi^{*}$ by performing the following update rule
\begin{equation}
    \label{eq: update_rule}
    Q_{t+1}(s,a) \leftarrow (1 - \alpha) Q_{t}(s,a) + \alpha [r + \gamma \max_{a^{'}}Q(s^{'},a^{'})],
\end{equation} 

where $\alpha$ is the learning rate and $r$ is the reward received upon taking action $a$ from state $s$. We usually derive a new policy $\pi^{'}$ which is greedy with respect to $Q^{\pi}(s,a)$, that is, ${\pi^{'}}$ $\in$ $argmax_{a} Q^{\pi}(s,a)$. The new policy $pi^{'}$ is guaranteed to be at least as good as policy $\pi$. Q-learning convergence and performance guarantees leading to an optimal policy $\pi^{*}$ is studied in \cite{watkins1992q, sutton2018reinforcement}. \\

\begin{figure}[h]
    \includegraphics[width=0.5\textwidth]{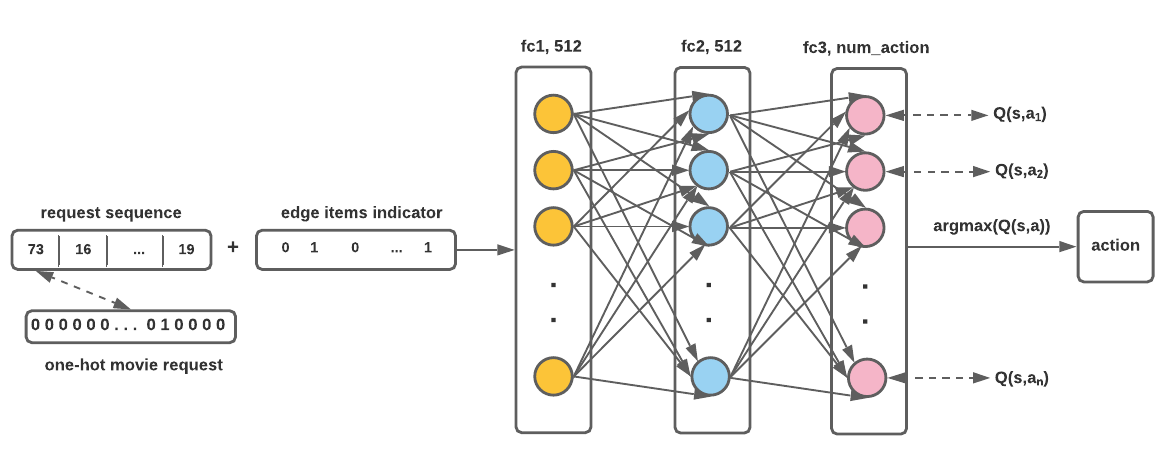}
	\caption{DeePref DQN Architecture} 
    \label{fig:DQN_architecture}
\end{figure}

Using DQN, we use a history of one-hot encoded user requests concatenated with a binary indicator vector representing edge items as input for DQN. Fig. \ref{fig:DQN_architecture} shows DeePref DQN architecture which consists of an input layer that outputs 512 neurons fully connected to an intermediate layer which outputs 512 neurons to the output layer which outputs the $Q(s,a)$ for each possible action $a$ being in state $s$. \\ 

\begin{figure}[h]
    \includegraphics[width=0.5\textwidth]{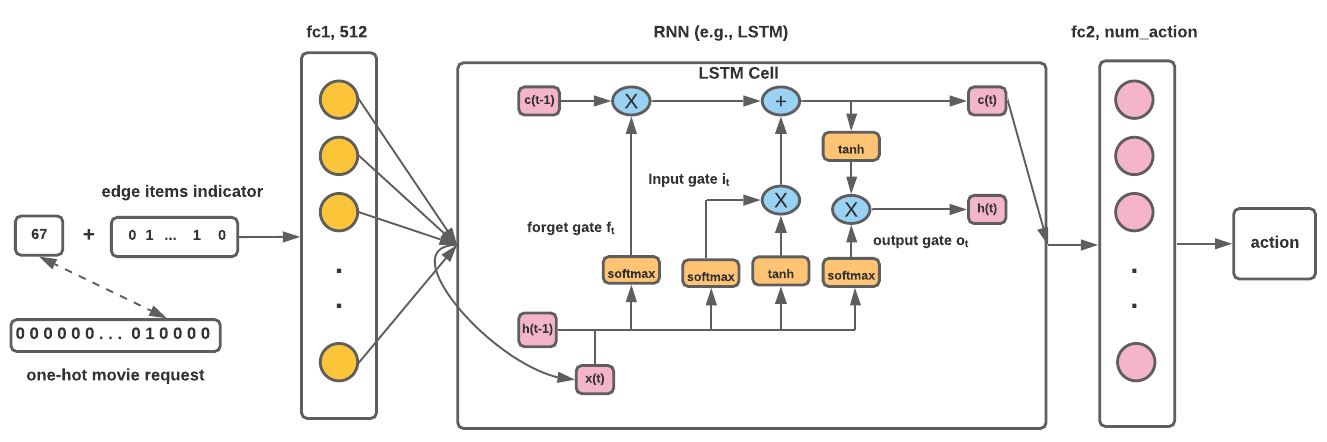}
	\caption{DeePref DRQN Architecture} 
    \label{fig:DRQN_architecture}
\end{figure}

\textsc{DQN} architecture does not capture the inter-sample dependencies across user requests since the input of DQN is a concatenation of user requests mimicking request history, which only captures spatial dependencies across samples. Therefore, we utilize Deep Recurrent Q-Network (DRQN)  which uses Recurrent Neural Networks (RNNs) such as Long Short-Term Memory (LSTM) \cite{hochreiter1997long} and Gated Recurrent Unit (GRU) \cite{cho2014learning} to counter the partial observability of the received user requests and to account for inter-sample dependencies across user requests which capture short-long term (temporal) dependencies of the previously seen samples (i.e., user requests). Thus, there is no need to encode user requests history as one-hot encoded as in the case of DQN. Instead, It suffices to only encode the currently requested content as one-hot encoded, since DRQN will be able to handle the inter-sample dependencies across user requests. Further, we append to the one hot encoded user request a binary indicator vector representing items at edge network. Afterward, the input is fed to a linear body as an input layer of size 512 followed by LSTM acting as RNN which is connected to an output layer where each neuron on the output layer represents the $Q(s,a)$ of each possible action $a$ being in state $s$. \textsc{DeePref DRQN} architecture is shown in Fig. \ref{fig:DRQN_architecture}. 

Moreover, \textsc{DRQNs} suffer from slowness during training when dealing with large sequences (i.e., large time-steps); therefore, we used Truncated Back Propagation Through Time (TBPTT) with $k_1 = k_2 = 300$ where $k_1$ is the number of forward-pass time-steps between LSTM updates, and $k_2$ is the number of time-steps to which we apply Back Propagation Through Time (BPTT). \\

\section{Experiment Design}
\label{sec:Experiment Design}

\begin{figure*}[h]
    \includegraphics[width=\textwidth]{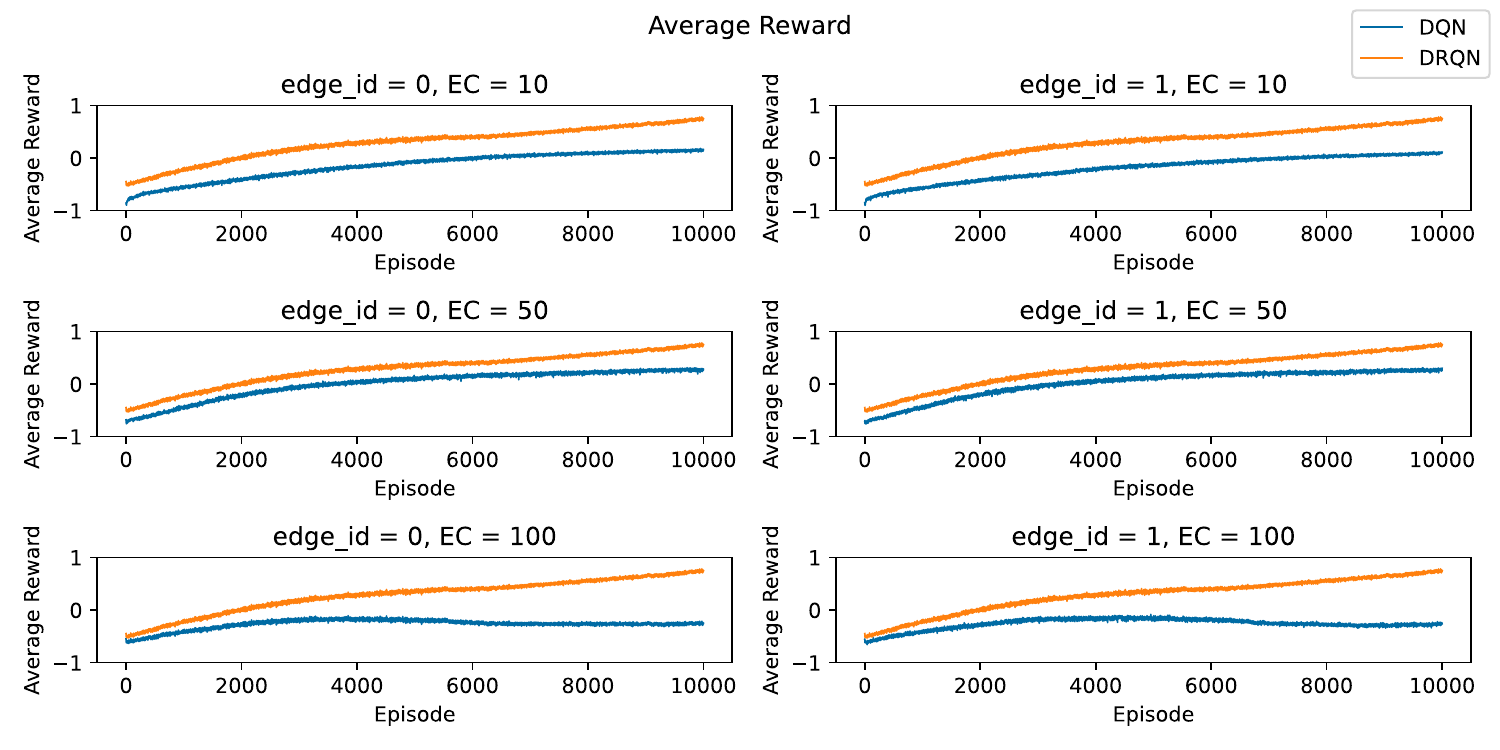}
	\caption{Average episodic reward during training phase} 
    \label{fig:average_reward_episodic}
\end{figure*}

\begin{table*}[t]
	\centering
\scalebox{0.7}{
\begin{tabular}{lllllllllllll}
		\toprule

Edge Capacity &                   EC = 10 &      &            &    &   EC = 50 &      &    & &   EC = 100 &      &            &       \\
\cmidrule(lr){1-13}

Metric       & Acc.    & Cov. & Timeliness & Aggr. & Acc.    & Cov. & Timeliness & Aggr. & Acc.     & Cov. & Timeliness & Aggr. \\

\cmidrule(lr){1-13}

Belady-Prefetch & 0.99       & 0.99    & $0.03\pm{0.19}$          & 0.99     & 0.99       & 0.99    & $20.39\pm{39.54}$          & 0.99     & 0.99        & 0.99   & $44.21\pm{60.79}$ & 0.99    \\



Top-k Popularity & 0.3       & 0.085    & $127.1\pm{61.74}$          & 0.05     & 0.5       & 0.27    & $108.64\pm{47.24}$          & 0.25     & 0.72        & 0.46   & $83.83\pm{24.97}$ & 0.5    \\

Top-k Size & 0.2       & 0.01    & $183.25\pm{35.55}$          & 0.05     & 0.22       & 0.05    & $175.35\pm{38.98}$          & 0.25     & 0.27        & 0.13   & $155.43\pm{43.07}$ & 0.5     \\

Popularity Recent & 0.1       & 0.1    & $8.28 \pm{3.25}$          &  1.0    & 0.26       & 0.26    & $28.18\pm{19.3}$          & 0.99     & 0.52        & 0.5   & $33.16\pm{23.77}$         & 0.97    \\

Popularity All    & 0.1       & 0.1    & $8.08\pm{3.39}$          & 0.97     & 0.23       & 0.23    & $31.59\pm{18.95}$          & 0.99     & 0.48        & 0.46   & $35.86\pm{27.48}$         & 0.96    \\

\addlinespace

\textbf{DeePref DQN}               & 0.42       & 0.15    & $98.0\pm{56.21}$          & 0.71     & 0.56       & 0.27    & $80.11\pm{47.37}$          & 0.73     & 0.76        & 0.56   & $66.81\pm{38.88}$         & 0.76    \\

\textbf{DeePref DRQN}              & 0.67       & 0.37    & $69\pm{54.56}$          & 0.5     & 0.76       & 0.58    & $83.3\pm{50.9}$          & 0.67     & 0.89        & 0.74   & $70.64\pm{38.37}$         & 0.82    \\


\end{tabular}
}
\caption{Testing Results on Edge ID = 1}
\label{tab:testing dataset}
\end{table*}

\subsection{Dataset}
\label{sec:Dataset}
\begin{figure}[h]
    \includegraphics[width=0.5\textwidth]{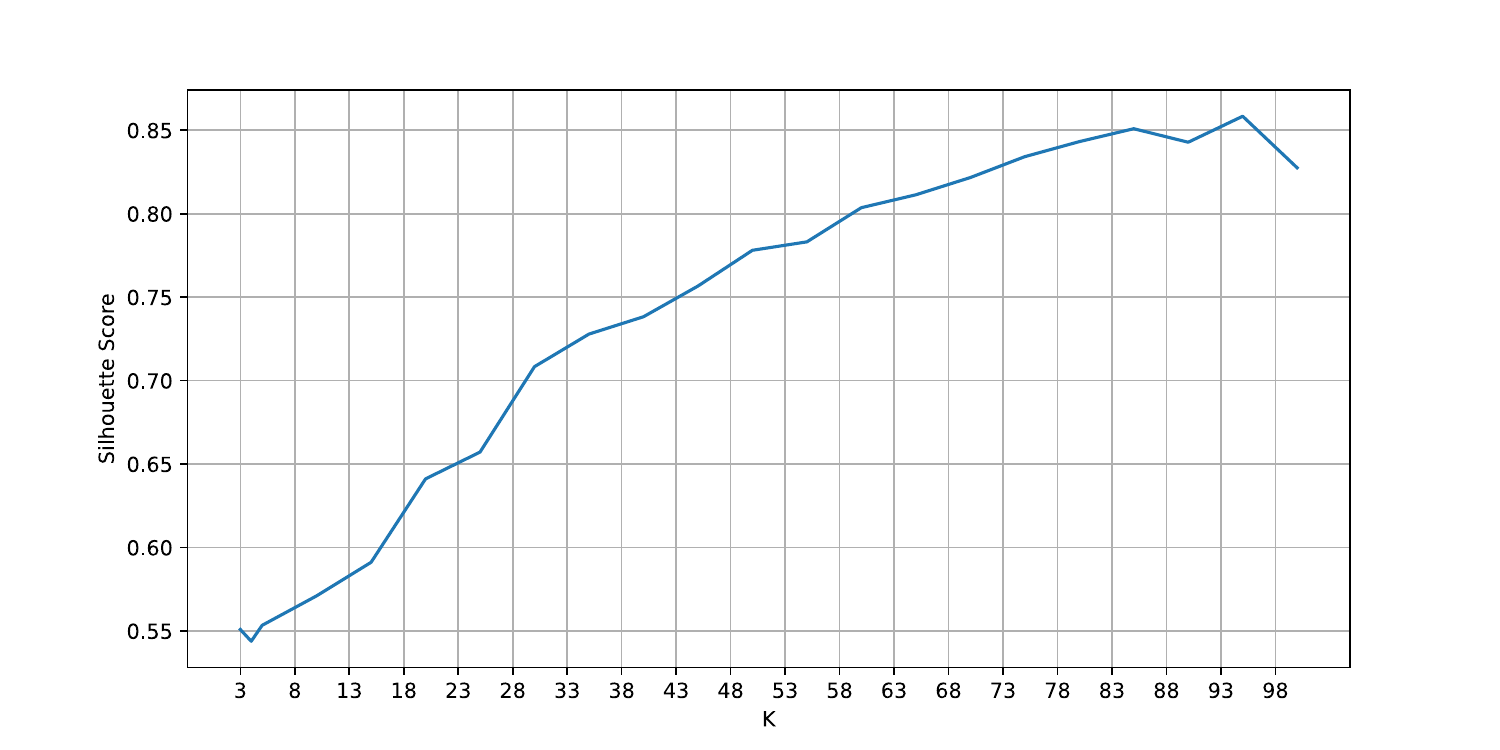}
	\caption{Silhouette Analysis to determine the number of k (edge nodes) in k-means clustering} 
    \label{fig:ksearch}
\end{figure}
We used the benchmark dataset MovieLens-100K (ml-100k) \cite{harper2016movielens} which is widely used in the research community to improve recommendation systems. MovieLens-100k (ml-100k) contains $100,000$ ratings from $943$ users on $1682$ movies. The data were collected over a period of seven months from September 19th, 1997 through April 22nd, 1998. ml-100k contains users' information such as age, gender, occupation, and zip code. We extract the longitude and latitude of each zip code from the US Census dataset and used k-means clustering to categorize user requests geographically into clusters, where each cluster represents an edge network. We used the elbow method with silhouette analysis to determine the number of clusters, as shown in Fig. \ref{fig:ksearch}. We chose $k=3$ as it has one of the lowest silhouette scores, as shown in Fig. \ref{fig:ksearch}. In other words, we select the number of agents to be 3 where we train two agents $(ID=0, ID=1)$ and we leave the third agent ($ID= 2$) for transfer learning. \\

We calculate the latency of each content to be prefetched from the cloud network based on the content size.  We assign content size randomly and to calculate the throughput $throughput = \frac{8 \times CWND}{RTT}$ (RFC 6349 \cite{constantine2011framework}). We assume Round Trip Time (RTT) to be $100$ ms and the receiver TCP window size (CWND) of $65536$ B. \\
\begin{figure*}[h]
    \includegraphics[width=\textwidth]{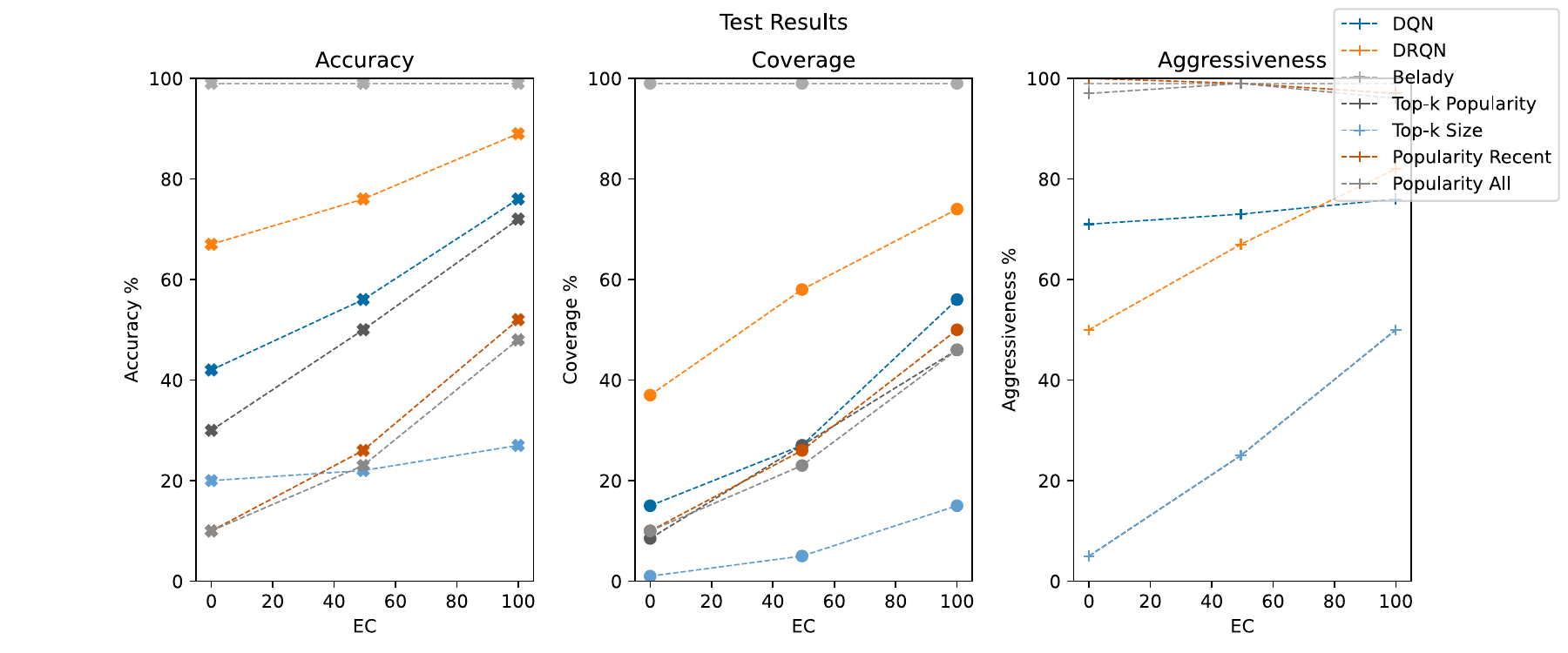}
	\caption{Test Results for Edge ID = 1} 
    \label{fig:metrics_vs_EC}
\end{figure*}
\begin{table*}[t]
	\centering
\scalebox{0.7}{
\begin{tabular}{lllllllllllll}
		\toprule

Edge Capacity &                   EC = 10 &      &            &    &   EC = 50 &      &    & &   EC = 100 &      &            &       \\

\cmidrule(lr){1-13}

Metric      & Acc.    & Cov. & Timeliness & Aggr. & Acc.    & Cov. & Timeliness & Aggr. & Acc.     & Cov. & Timeliness & Aggr. \\
\cmidrule(lr){1-13}

Belady-Prefetch & 0.99       & 0.99    & $0.04\pm{0.21}$          & 0.99     & 0.99       & 0.99    & $0.17\pm{0.37}$          & 0.89     & 0.99        & 0.99   & $0.27\pm{0.44}$ & 0.99    \\



Top-k Popularity & 0.23       & 0.05    & $150.2\pm{50.95}$          & 0.05     & 0.36       & 0.2    & $131.76\pm{41.67}$          & 0.25     & 0.42        & 0.34   & $108.99\pm{32.86}$ & 0.5    \\

Top-k Size & 0.0       & 0.0    & $200\pm{0.0}$          & 0.05     & 0.28       & 0.07    & $160.53\pm{53.99}$          & 0.25     & 0.29        & 0.15   & $150.19\pm{43.96}$ & 0.5    \\

Popularity Recent & 0.12       & 0.12    & $7.48\pm{3.39}$          & 0.95     & 0.23       & 0.25    & $29.52\pm{19.53}$          & 0.99     & 0.43        & 0.51   & $34.55\pm{27.46}$ & 0.93    \\

Popularity All    & 0.08       & 0.07    & $8.67\pm{3.92}$          & 0.85     & 0.21       & 0.22    & $31.41\pm{19.0}$          & 0.99     & 0.41        & 0.48   & $36.31\pm{28.87}$         & 0.93    \\

\addlinespace

\textbf{DeePref DQN}               & 0.32       & 0.2    & $102.92\pm{66.88}$          & 0.23     & 0.42       & 0.37    & $84.54\pm{50.04}$          & 0.44     & 0.65        & 0.46   & $70.27\pm{40.36}$      & 0.32    \\

\textbf{DeePref DRQN}              & 0.56       & 0.34    & $85.32\pm{49.59}$          & 0.25     & 0.66       & 0.48    & $74.55\pm{47.15}$          & 0.42     & 0.72        & 0.58   & $72\pm{50.36}$         & 0.23    \\


\end{tabular}
}
\caption{Transfer Results of Edge ID = 1 to Edge ID = 2}
\label{tab:transfer dataset}
\end{table*}
\subsection{Results}
\label{sec:results}
We trained two agents $(ID=0, ID=1)$ using \textsc{DeePref DQN} and \textsc{DeePRef DRQN} architectures and compared our testing and transferring results to the following baseline prefetching algorithms:
\begin{itemize}
    \item \textbf{\emph{Belady-Prefetch}}: The edge network will prefetch (fetch) the item that is currently being requested in case of a cache miss. Once the item is received from the cloud network, Belady eviction algorithm \cite{belady1966study} will be used to evict the item in which its occurrence happens the furthest. This approach will serve as an oracle in our experiments. \\

    

    \item \textbf{\emph{Top-k Popularity}}: The edge network will prefetch Top-k popular video content that is known to the cloud network. The cloud network periodically provides a list of top-k popular video content to edge networks based on their geographical location \cite{alkassab2017benefits}. \\

    \item \textbf{\emph{Top-k Size}}: The edge network will prefetch Top-k largest video content that is known to the cloud network. The cloud network periodically provides a list of top-k largest video content to edge networks based on their geographical location \cite{alkassab2017benefits}. \\

    \item \textbf{\emph{Popularity Recent}}: Using on-edge prefetching, the agent (i.e., edge network) will prefetch the most popular content using a sliding time window of 24 hours. If the most popular item seen in the past 24 hours already exists at the edge's storage, then no item will be prefetched. \\
    
    \item \textbf{\emph{Popularity All}}: Using on-edge prefetching, the agent (i.e., edge network) will prefetch the most popular item that has been seen so far since the beginning of the experiment. \\

\end{itemize}

We use the following evaluation metrics to compare DeePref DQN and DRQN architectures with baselines: 
\begin{itemize}
    \item \textbf{\emph{Prefetching Accuracy (Precision Ratio)}}: Prefetching Accuracy describes how precise the prefetching decisions are in terms of sent prefetches. Prefetching accuracy is also referred to as the precision ratio of prefetching, which is defined below:
    \begin{equation}
    \label{eq: Prefetching Accuracy}
    Accuracy\:(Precision) = \frac{\#hits}{\#hits + \#misses} = \frac{\#hits}{\#user\: requests}
    \end{equation}    
    \item \textbf{\emph{Prefetching Coverage (Recall Ratio)}}: Prefetching Coverage describes the usefulness of sent prefetches. Prefetching coverage is also referred to as recall ratio, which is defined in the following equation:
    \begin{equation}
    \label{eq: Prefetching Coverage}
    Coverage\:(Recall) = \frac{\#used\:prefetches}{\#sent\:prefetches}
    \end{equation}   

    \item \textbf{\emph{Prefetching Aggressiveness}}: Prefetching Aggressiveness describes how many prefetches had been sent throughout the entire experiment. This metric is used to compare prefetching algorithms when they are achieving similar prefetching accuracy and coverage and is defined below:
    \begin{equation}
    \label{eq: Prefetching Aggressiveness}
    Aggressiveness= \frac{\#sent\:prefetches}{\#hits + \#misses} = \frac{\#sent\:prefetches}{\#user\:requests}
    \end{equation}     

    \item \textbf{\emph{Prefetching Timeliness}}: Timeliness measures how long the item had been residing at edge storage before it got evicted. If the item received a cache hit, its timeliness timer is refreshed. This metric in correlation with prefetching aggressiveness is important since we are decoupling prefetching algorithms from eviction algorithms, as stated previously.
\end{itemize}

We trained two edge networks using \textsc{DeePref DQN} and \textsc{DeePref DRQN} models. Both trained edge networks, namely: $Edge ID = 0$ and $Edge ID = 1$ show similar results. We highlight the results of only $Edge ID = 1$ for brevity. As shown in Fig. \ref{fig:average_reward_episodic}, \textsc{DeePref DRQN} model clearly outperforms \textsc{DeePref DQN} model during the training phase in terms of average episodic reward. This can be explained due the fact that the DRQN agent is equipped with a replay memory in the form of Recurrent Neural Networks (RNNs) such as LSTMs and GRUs which capture the spatio-temporal dependencies across input samples (i.e., user requests). We design the replay memory of \textsc{DeePref DRQN} using both LSTMs and GRUs where both show comparative results. Moreover, the replay memory is random instead of sequential since user requests are coming from heterogeneous networks that are in nature represent a non i.i.d distribution.  \\

More importantly, we address and discuss how well \textsc{DeePref DQN} and \textsc{DeePref DRQN} models perform on unseen scenarios, namely: 1) testing dataset which consists of unseen user requests drawn from the same distribution of user requests that the edge network serves, and 2) transfer dataset which consists of unseen user requests obtained from a different edge network cluster, that is $Edge ID = 2$ since possible transfer learning of statistical model such as \textsc{DeePref DQN} and \textsc{DeePref DRQN} can occur across edge networks. \\

As shown in Table \ref{tab:testing dataset} and depicted in Fig. \ref{fig:metrics_vs_EC}, the testing results show that both \textsc{DeePref DQN} and \textsc{DeePref DRQN} architectures outperform traditional prefetching schemes such as \textit{Top-k Popularity} and \textit{Popularity Recent} in terms of prefetching accuracy and prefetching coverage. \textsc{DeePref DRQN} outperforms baseline approaches with an increase of at least [$17\%$, $28\%$] in terms of prefetching accuracy and coverage, respectively. As presented in Table \ref{tab:testing dataset}, our results show that more aggressive algorithms tend to have better prefetching accuracy and coverage since some prefetching algorithms such as \textit{Top-k Size} may decide not to prefetch at all, missing $100\%$ of the chances of observing a prefetch hit. Furthermore, Fig.\ref{fig:metrics_vs_EC} shows that the increase in the Edge Capacity (EC) increases both prefetching accuracy and coverage. \\

Furthermore, when transferring \textsc{DeePref DRQN} model from the trained edge (e.g., $Edge ID =1$) to an entirely new edge network (e.g., $Edge ID =2$) which represents a different statistical distribution, our results show similar improvements compared to baseline approaches. As shown in Table \ref{tab:transfer dataset}, \textsc{DeePref DRQN} improves both prefetching accuracy and prefetching coverage compared to baseline approaches with an increase of [$30\%$,$10\%$], respectively. \\ 



\section{Concluding Remarks}
\label{sec:Concluding Remarks}
We proposed \textsc{DeePref DRQN}, a prefetcher for video prefetching in CDNs. \textsc{DeePref} is a Deep Reinforcement Learning agent that utilizes the network latency in terms of punishments and rewards to make prefetching decisions online. \textsc{DeePref} is a prefetcher that is implemented at edge networks and is agnostic to hardware design, operating system, and application. Our results show that \textsc{DeePref} DRQN outperforms baseline approaches in terms of prefetching accuracy and prefetching coverage. Our results also show that aggressive approaches tend to have better prefetching accuracy and coverage. \textsc{DeePref} is an auto-aggressive prefetcher in the sense that it adapts to user requests online without any changes in the model architecture. \\
\section*{Acknowledgement}
\label{sec:Acknowledgement}
We thank the Research Computing team at the University of South Carolina for their efforts in making High-Performance Computing (HPC) clusters available for us to run our experiments. \\

\bibliographystyle{plain}
\bibliography{main}

\begin{thebibliography}{10}

\bibitem{adhikari2014measurement}
Vijay~K Adhikari, Yang Guo, Fang Hao, Volker Hilt, Zhi-Li Zhang, Matteo
  Varvello, and Moritz Steiner.
\newblock Measurement study of netflix, hulu, and a tale of three cdns.
\newblock {\em IEEE/ACM Transactions on Networking}, 23(6):1984--1997, 2014.

\bibitem{aguilar2023space}
Jes{\'u}s Aguilar-Armijo, Christian Timmerer, and Hermann Hellwagner.
\newblock Space: Segment prefetching and caching at the edge for adaptive video
  streaming.
\newblock {\em IEEE Access}, 11:21783--21798, 2023.

\bibitem{al2020effectively}
Hasan Al~Maruf and Mosharaf Chowdhury.
\newblock Effectively prefetching remote memory with leap.
\newblock In {\em 2020 USENIX Annual Technical Conference (USENIX ATC 20)},
  pages 843--857, 2020.

\bibitem{albericio2012abs}
Jorge Albericio, Rub{\'e}n Gran, Pablo Ib{\'a}nez, V{\'\i}ctor Vi{\~n}als, and
  Jose~Mar{\'\i}a Llaber{\'\i}a.
\newblock Abs: A low-cost adaptive controller for prefetching in a banked
  shared last-level cache.
\newblock {\em ACM Transactions on Architecture and Code Optimization (TACO)},
  8(4):1--20, 2012.

\bibitem{alkassab2017benefits}
Nawras Alkassab, Chin-Tser Huang, Yu~Chen, Baek-Young Choi, and Sejun Song.
\newblock Benefits and schemes of prefetching from cloud to fog networks.
\newblock In {\em 2017 IEEE 6th International Conference on Cloud Networking
  (CloudNet)}, pages 1--5. IEEE, 2017.

\bibitem{bartels1999potentials}
Gretta Bartels, Anna Karlin, Darrell Anderson, Jeffrey Chase, Henry Levy, and
  Geoffrey Voelker.
\newblock Potentials and limitations of fault-based markov prefetching for
  virtual memory pages.
\newblock {\em ACM SIGMETRICS Performance Evaluation Review}, 27(1):206--207,
  1999.

\bibitem{behravesh2022machine}
Rasoul Behravesh, Akhila Rao, Daniel~F Perez-Ramirez, Davit Harutyunyan,
  Roberto Riggio, and Magnus Boman.
\newblock Machine learning at the mobile edge: The case of dynamic adaptive
  streaming over http (dash).
\newblock {\em IEEE Transactions on Network and Service Management},
  19(4):4779--4793, 2022.

\bibitem{belady1966study}
Laszlo~A. Belady.
\newblock A study of replacement algorithms for a virtual-storage computer.
\newblock {\em IBM Systems journal}, 5(2):78--101, 1966.

\bibitem{bera2021pythia}
Rahul Bera, Konstantinos Kanellopoulos, Anant Nori, Taha Shahroodi, Sreenivas
  Subramoney, and Onur Mutlu.
\newblock Pythia: A customizable hardware prefetching framework using online
  reinforcement learning.
\newblock In {\em MICRO-54: 54th Annual IEEE/ACM International Symposium on
  Microarchitecture}, pages 1121--1137, 2021.

\bibitem{breslau1999web}
Lee Breslau, Pei Cao, Li~Fan, Graham Phillips, and Scott Shenker.
\newblock Web caching and zipf-like distributions: Evidence and implications.
\newblock In {\em IEEE INFOCOM'99. Conference on Computer Communications.
  Proceedings. Eighteenth Annual Joint Conference of the IEEE Computer and
  Communications Societies. The Future is Now (Cat. No. 99CH36320)}, volume~1,
  pages 126--134. IEEE, 1999.

\bibitem{cao1995study}
Pei Cao, Edward~W Felten, Anna~R Karlin, and Kai Li.
\newblock A study of integrated prefetching and caching strategies.
\newblock {\em ACM SIGMETRICS Performance Evaluation Review}, 23(1):188--197,
  1995.

\bibitem{chang1999automatic}
Fay Chang and Garth Gibson.
\newblock Automatic i/o hint generation through speculative execution.
\newblock USENIX, 1999.

\bibitem{chen1995effective}
Tien-Fu Chen and Jean-Loup Baer.
\newblock Effective hardware-based data prefetching for high-performance
  processors.
\newblock {\em IEEE transactions on computers}, 44(5):609--623, 1995.

\bibitem{chiang2015adaptive}
Ron~C Chiang, Ahsen~J Uppal, and H~Howie Huang.
\newblock An adaptive io prefetching approach for virtualized data centers.
\newblock {\em IEEE Transactions on Services Computing}, 10(3):328--340, 2015.

\bibitem{cho2014learning}
Kyunghyun Cho, Bart Van~Merri{\"e}nboer, Caglar Gulcehre, Dzmitry Bahdanau,
  Fethi Bougares, Holger Schwenk, and Yoshua Bengio.
\newblock Learning phrase representations using rnn encoder-decoder for
  statistical machine translation.
\newblock {\em arXiv preprint arXiv:1406.1078}, 2014.

\bibitem{cisco2018cisco}
VNI Cisco.
\newblock Cisco visual networking index: Forecast and trends, 2017--2022.
\newblock {\em White Paper}, 2018.

\bibitem{constantine2011framework}
Barry Constantine, Gilles Forget, Ruediger Geib, and Reinhard Schrage.
\newblock Framework for tcp throughput testing.
\newblock Technical report, 2011.

\bibitem{cranley2006user}
Nicola Cranley, Philip Perry, and Liam Murphy.
\newblock User perception of adapting video quality.
\newblock {\em International Journal of Human-Computer Studies},
  64(8):637--647, 2006.

\bibitem{ferdman2011proactive}
Michael Ferdman, Cansu Kaynak, and Babak Falsafi.
\newblock Proactive instruction fetch.
\newblock In {\em Proceedings of the 44th Annual IEEE/ACM International
  Symposium on Microarchitecture}, pages 152--162, 2011.

\bibitem{ge2017toward}
Chang Ge, Ning Wang, Gerry Foster, and Mick Wilson.
\newblock Toward qoe-assured 4k video-on-demand delivery through mobile edge
  virtualization with adaptive prefetching.
\newblock {\em IEEE Transactions on Multimedia}, 19(10):2222--2237, 2017.

\bibitem{gill2007amp}
Binny~S Gill and Luis Angel~D Bathen.
\newblock Amp: Adaptive multi-stream prefetching in a shared cache.
\newblock In {\em FAST}, volume~7, pages 185--198, 2007.

\bibitem{gill2005sarc}
Binny~S Gill and Dharmendra~S Modha.
\newblock Sarc: Sequential prefetching in adaptive replacement cache.
\newblock In {\em USENIX Annual Technical Conference, General Track}, pages
  293--308, 2005.

\bibitem{gill2007youtube}
Phillipa Gill, Martin Arlitt, Zongpeng Li, and Anirban Mahanti.
\newblock Youtube traffic characterization: a view from the edge.
\newblock In {\em Proceedings of the 7th ACM SIGCOMM conference on Internet
  measurement}, pages 15--28, 2007.

\bibitem{guo2007compiler}
Yao Guo.
\newblock {\em Compiler-assisted hardware-based data prefetching for next
  generation processors}.
\newblock University of Massachusetts Amherst, 2007.

\bibitem{harper2016movielens}
F~Maxwell Harper and Joseph~A Konstan.
\newblock The movielens datasets: History and context.
\newblock {\em Acm transactions on interactive intelligent systems (tiis)},
  5(4):19, 2016.

\bibitem{hochreiter1997long}
Sepp Hochreiter and J{\"u}rgen Schmidhuber.
\newblock Long short-term memory.
\newblock {\em Neural computation}, 9(8):1735--1780, 1997.

\bibitem{hu2017musa}
Wen Hu, Jiahui Huang, Zhi Wang, Peng Wang, Kun Yi, Yonggang Wen, Kaiyan Chu,
  and Lifeng Sun.
\newblock Musa: Wi-fi ap-assisted video prefetching via tensor learning.
\newblock In {\em 2017 IEEE/ACM 25th International Symposium on Quality of
  Service (IWQoS)}, pages 1--6. IEEE, 2017.

\bibitem{jiang2012improving}
Junchen Jiang, Vyas Sekar, and Hui Zhang.
\newblock Improving fairness, efficiency, and stability in http-based adaptive
  video streaming with festive.
\newblock In {\em Proceedings of the 8th international conference on Emerging
  networking experiments and technologies}, pages 97--108, 2012.

\bibitem{jiang2005dulo}
Song Jiang, Xiaoning Ding, Feng Chen, Enhua Tan, and Xiaodong Zhang.
\newblock Dulo: an effective buffer cache management scheme to exploit both
  temporal and spatial locality.
\newblock In {\em Proceedings of the 4th conference on USENIX Conference on
  File and Storage Technologies}, volume~4, pages 8--8, 2005.

\bibitem{jiang2013prefetching}
Song Jiang, Xiaoning Ding, Yuehai Xu, and Kei Davis.
\newblock A prefetching scheme exploiting both data layout and access history
  on disk.
\newblock {\em ACM Transactions on Storage (TOS)}, 9(3):1--23, 2013.

\bibitem{joo2011fast}
Yongsoo Joo, Junhee Ryu, Sangsoo Park, and Kang~G Shin.
\newblock $\{$FAST$\}$: Quick application launch on $\{$Solid-State$\}$ drives.
\newblock In {\em 9th USENIX Conference on File and Storage Technologies (FAST
  11)}, 2011.

\bibitem{kaplan2010users}
Andreas~M Kaplan and Michael Haenlein.
\newblock Users of the world, unite! the challenges and opportunities of social
  media.
\newblock {\em Business horizons}, 53(1):59--68, 2010.

\bibitem{kimbrel2000near}
Tracy Kimbrel and Anna~R Karlin.
\newblock Near-optimal parallel prefetching and caching.
\newblock {\em SIAM Journal on computing}, 29(4):1051--1082, 2000.

\bibitem{kober2013reinforcement}
Jens Kober, J~Andrew Bagnell, and Jan Peters.
\newblock Reinforcement learning in robotics: A survey.
\newblock {\em The International Journal of Robotics Research},
  32(11):1238--1274, 2013.

\bibitem{kolli2013rdip}
Aasheesh Kolli, Ali Saidi, and Thomas~F Wenisch.
\newblock Rdip: Return-address-stack directed instruction prefetching.
\newblock In {\em 2013 46th Annual IEEE/ACM International Symposium on
  Microarchitecture (MICRO)}, pages 260--271. IEEE, 2013.

\bibitem{langley2010transport}
Adam Langley et~al.
\newblock Transport layer security (tls) snap start.
\newblock {\em Working Draft}, 2010.

\bibitem{langley2016transport}
Adam Langley, Nagendra Modadugu, and Bodo Moeller.
\newblock Transport layer security (tls) false start.
\newblock Technical report, 2016.

\bibitem{lattner2005automatic}
Chris Lattner and Vikram Adve.
\newblock Automatic pool allocation: improving performance by controlling data
  structure layout in the heap.
\newblock {\em ACM Sigplan Notices}, 40(6):129--142, 2005.

\bibitem{lecun2015deep}
Yann LeCun, Yoshua Bengio, and Geoffrey Hinton.
\newblock Deep learning.
\newblock {\em nature}, 521(7553):436--444, 2015.

\bibitem{levine2016end}
Sergey Levine, Chelsea Finn, Trevor Darrell, and Pieter Abbeel.
\newblock End-to-end training of deep visuomotor policies.
\newblock {\em The Journal of Machine Learning Research}, 17(1):1334--1373,
  2016.

\bibitem{liang2015integrated}
Ke~Liang, Jia Hao, Roger Zimmermann, and David~KY Yau.
\newblock Integrated prefetching and caching for adaptive video streaming over
  http: an online approach.
\newblock In {\em Proceedings of the 6th ACM Multimedia Systems Conference},
  pages 142--152. ACM, 2015.

\bibitem{mcafee2012big}
Andrew McAfee, Erik Brynjolfsson, Thomas~H Davenport, DJ~Patil, and Dominic
  Barton.
\newblock Big data: the management revolution.
\newblock {\em Harvard business review}, 90(10):60--68, 2012.

\bibitem{mell2011nist}
Peter Mell, Tim Grance, et~al.
\newblock The nist definition of cloud computing.
\newblock 2011.

\bibitem{mittal2016survey}
Sparsh Mittal.
\newblock A survey of recent prefetching techniques for processor caches.
\newblock {\em ACM Computing Surveys (CSUR)}, 49(2):1--35, 2016.

\bibitem{mnih2013playing}
Volodymyr Mnih, Koray Kavukcuoglu, David Silver, Alex Graves, Ioannis
  Antonoglou, Daan Wierstra, and Martin Riedmiller.
\newblock Playing atari with deep reinforcement learning.
\newblock {\em arXiv preprint arXiv:1312.5602}, 2013.

\bibitem{pallis2006insight}
George Pallis and Athena Vakali.
\newblock Insight and perspectives for content delivery networks.
\newblock {\em Communications of the ACM}, 49(1):101--106, 2006.

\bibitem{panda2016spac}
Biswabandan Panda.
\newblock Spac: A synergistic prefetcher aggressiveness controller for
  multi-core systems.
\newblock {\em IEEE Transactions on Computers}, 65(12):3740--3753, 2016.

\bibitem{panda2015caffeine}
Biswabandan Panda and Shankar Balachandran.
\newblock Caffeine: A utility-driven prefetcher aggressiveness engine for
  multicores.
\newblock {\em ACM Transactions on Architecture and Code Optimization (TACO)},
  12(3):1--25, 2015.

\bibitem{pathan2007taxonomy}
Al-Mukaddim~Khan Pathan and Rajkumar Buyya.
\newblock A taxonomy and survey of content delivery networks.
\newblock {\em Grid Computing and Distributed Systems Laboratory, University of
  Melbourne, Technical Report}, 4:70, 2007.

\bibitem{patterson1995informed}
R~Hugo Patterson, Garth~A Gibson, Eka Ginting, Daniel Stodolsky, and Jim
  Zelenka.
\newblock Informed prefetching and caching.
\newblock In {\em Proceedings of the fifteenth ACM symposium on Operating
  systems principles}, pages 79--95, 1995.

\bibitem{peled2015semantic}
Leeor Peled, Shie Mannor, Uri Weiser, and Yoav Etsion.
\newblock Semantic locality and context-based prefetching using reinforcement
  learning.
\newblock In {\em 2015 ACM/IEEE 42nd Annual International Symposium on Computer
  Architecture (ISCA)}, pages 285--297. IEEE, 2015.

\bibitem{puterman2014markov}
Martin~L Puterman.
\newblock {\em Markov decision processes: discrete stochastic dynamic
  programming}.
\newblock John Wiley \& Sons, 2014.

\bibitem{rabbah2004compiler}
Rodric~M Rabbah, Hariharan Sandanagobalane, Mongkol Ekpanyapong, and Weng-Fai
  Wong.
\newblock Compiler orchestrated prefetching via speculation and predication.
\newblock {\em ACM SIGARCH Computer Architecture News}, 32(5):189--198, 2004.

\bibitem{rimal2009taxonomy}
Bhaskar~Prasad Rimal, Eunmi Choi, and Ian Lumb.
\newblock A taxonomy and survey of cloud computing systems.
\newblock In {\em 2009 Fifth International Joint Conference on INC, IMS and
  IDC}, pages 44--51. Ieee, 2009.

\bibitem{russel2013artificial}
Stuart Russel, Peter Norvig, et~al.
\newblock {\em Artificial intelligence: a modern approach}.
\newblock Pearson Education Limited, 2013.

\bibitem{shacham2002fast}
Hovav Shacham, Dan Boneh, et~al.
\newblock Fast-track session establishment for tls.
\newblock In {\em NDSS}. Citeseer, 2002.

\bibitem{shafiq2014revisiting}
Muhammad~Zubair Shafiq, Alex~X Liu, and Amir~R Khakpour.
\newblock Revisiting caching in content delivery networks.
\newblock In {\em The 2014 ACM international conference on Measurement and
  modeling of computer systems}, pages 567--568, 2014.

\bibitem{sham2021intelligent}
Eht~E Sham and Deo~Prakash Vidyarthi.
\newblock Intelligent admission control manager for fog-integrated cloud: A
  hybrid machine learning approach.
\newblock {\em Concurrency and Computation: Practice and Experience}, page
  e6687, 2021.

\bibitem{shi2016edge}
Weisong Shi, Jie Cao, Quan Zhang, Youhuizi Li, and Lanyu Xu.
\newblock Edge computing: Vision and challenges.
\newblock {\em IEEE Internet of Things Journal}, 3(5):637--646, 2016.

\bibitem{silver2016mastering}
David Silver, Aja Huang, Chris~J Maddison, Arthur Guez, Laurent Sifre, George
  Van Den~Driessche, Julian Schrittwieser, Ioannis Antonoglou, Veda
  Panneershelvam, Marc Lanctot, et~al.
\newblock Mastering the game of go with deep neural networks and tree search.
\newblock {\em nature}, 529(7587):484--489, 2016.

\bibitem{soundararajan2008context}
Gokul Soundararajan, Madalin Mihailescu, and Cristiana Amza.
\newblock $\{$Context-Aware$\}$ prefetching at the storage server.
\newblock In {\em 2008 USENIX Annual Technical Conference (USENIX ATC 08)},
  2008.

\bibitem{stark2012case}
Emily Stark, Lin-Shung Huang, Dinesh Israni, Collin Jackson, and Dan Boneh.
\newblock The case for prefetching and prevalidating tls server certificates.
\newblock In {\em NDSS}, volume~12, 2012.

\bibitem{summers2016characterizing}
Jim Summers, Tim Brecht, Derek Eager, and Alex Gutarin.
\newblock Characterizing the workload of a netflix streaming video server.
\newblock In {\em 2016 IEEE International Symposium on Workload
  Characterization (IISWC)}, pages 1--12. IEEE, 2016.

\bibitem{summers2014automated}
Jim Summers, Tim Brecht, Derek Eager, Tyler Szepesi, Ben Cassell, and Bernard
  Wong.
\newblock Automated control of aggressive prefetching for http streaming video
  servers.
\newblock In {\em Proceedings of International Conference on Systems and
  Storage}, pages 1--11, 2014.

\bibitem{summers2012methodologies}
Jim Summers, Tim Brecht, Derek Eager, and Bernard Wong.
\newblock Methodologies for generating http streaming video workloads to
  evaluate web server performance.
\newblock In {\em Proceedings of the 5th Annual International Systems and
  Storage Conference}, pages 1--12, 2012.

\bibitem{sun2019combining}
Gongjin Sun, Junjie Shen, and Alexander~V Veidenbaum.
\newblock Combining prefetch control and cache partitioning to improve
  multicore performance.
\newblock In {\em 2019 IEEE International Parallel and Distributed Processing
  Symposium (IPDPS)}, pages 953--962. IEEE, 2019.

\bibitem{sutton2018reinforcement}
Richard~S Sutton and Andrew~G Barto.
\newblock {\em Reinforcement learning: An introduction}.
\newblock MIT press, 2018.

\bibitem{sutton1998introduction}
Richard~S Sutton, Andrew~G Barto, et~al.
\newblock {\em Introduction to reinforcement learning}, volume~2.
\newblock MIT press Cambridge, 1998.

\bibitem{sutton1999policy}
Richard~S Sutton, David McAllester, Satinder Singh, and Yishay Mansour.
\newblock Policy gradient methods for reinforcement learning with function
  approximation.
\newblock {\em Advances in neural information processing systems},
  12:1057--1063, 1999.

\bibitem{tesauro1995temporal}
Gerald Tesauro.
\newblock Temporal difference learning and td-gammon.
\newblock {\em Communications of the ACM}, 38(3):58--68, 1995.

\bibitem{tsitsiklis1997analysis}
John~N Tsitsiklis and Benjamin Van~Roy.
\newblock Analysis of temporal-diffference learning with function
  approximation.
\newblock In {\em Advances in neural information processing systems}, pages
  1075--1081, 1997.

\bibitem{vandebogart2009reducing}
Steve VanDeBogart, Christopher Frost, and Eddie Kohler.
\newblock Reducing seek overhead with application-directed prefetching.
\newblock In {\em USENIX Annual Technical Conference}, 2009.

\bibitem{vander1997caches}
Steven~P Vander~Wiel and David~J Lilja.
\newblock When caches aren't enough: Data prefetching techniques.
\newblock {\em Computer}, 30(7):23--30, 1997.

\bibitem{vander1999compiler}
Steven~P Vander~Wiel and David~J Lilja.
\newblock A compiler-assisted data prefetch controller.
\newblock In {\em Proceedings 1999 IEEE International Conference on Computer
  Design: VLSI in Computers and Processors (Cat. No. 99CB37040)}, pages
  372--377. IEEE, 1999.

\bibitem{venkataramani2002potential}
Arun Venkataramani, Praveen Yalagandula, Ravindranath Kokku, Sadia Sharif, and
  Mike Dahlin.
\newblock The potential costs and benefits of long-term prefetching for content
  distribution.
\newblock {\em Computer Communications}, 25(4):367--375, 2002.

\bibitem{voelker1998implementing}
Geoffrey~M Voelker, Eric~J Anderson, Tracy Kimbrel, Michael~J Feeley, Jeffrey~S
  Chase, Anna~R Karlin, and Henry~M Levy.
\newblock Implementing cooperative prefetching and caching in a
  globally-managed memory system.
\newblock In {\em Proceedings of the 1998 ACM SIGMETRICS joint international
  conference on Measurement and modeling of computer systems}, pages 33--43,
  1998.

\bibitem{waldspurger2015efficient}
Carl~A Waldspurger, Nohhyun Park, Alexander Garthwaite, and Irfan Ahmad.
\newblock Efficient $\{$MRC$\}$ construction with $\{$SHARDS$\}$.
\newblock In {\em 13th USENIX Conference on File and Storage Technologies (FAST
  15)}, pages 95--110, 2015.

\bibitem{watkins1992q}
Christopher~JCH Watkins and Peter Dayan.
\newblock Q-learning.
\newblock {\em Machine learning}, 8(3-4):279--292, 1992.

\bibitem{williams1992simple}
Ronald~J Williams.
\newblock Simple statistical gradient-following algorithms for connectionist
  reinforcement learning.
\newblock {\em Machine learning}, 8(3-4):229--256, 1992.

\bibitem{wu2021optimization}
Hao Wu, Youlong Luo, and Chunlin Li.
\newblock Optimization of heat-based cache replacement in edge computing
  system.
\newblock {\em The Journal of Supercomputing}, 77:2268--2301, 2021.

\bibitem{yang2017mithril}
Juncheng Yang, Reza Karimi, Trausti S{\ae}mundsson, Avani Wildani, and Ymir
  Vigfusson.
\newblock Mithril: mining sporadic associations for cache prefetching.
\newblock In {\em Proceedings of the 2017 Symposium on Cloud Computing}, pages
  66--79, 2017.

\bibitem{yang2016tombolo}
Suli Yang, Kiran Srinivasan, Kishore Udayashankar, Swetha Krishnan, Jingxin
  Feng, Yupu Zhang, Andrea~C Arpaci-Dusseau, and Remzi~H Arpaci-Dusseau.
\newblock Tombolo: Performance enhancements for cloud storage gateways.
\newblock In {\em 2016 32nd Symposium on Mass Storage Systems and Technologies
  (MSST)}, pages 1--14. IEEE, 2016.

\bibitem{zhu2020ctdgm}
Dongjie Zhu, Haiwen Du, Yundong Sun, and Zhaoshuo Tian.
\newblock Ctdgm: A data grouping model based on cache transaction for
  unstructured data storage systems.
\newblock {\em arXiv preprint arXiv:2009.14414}, 2020.

\bibitem{zink2009characteristics}
Michael Zink, Kyoungwon Suh, Yu~Gu, and Jim Kurose.
\newblock Characteristics of youtube network traffic at a campus
  network--measurements, models, and implications.
\newblock {\em Computer networks}, 53(4):501--514, 2009.

\bibitem{zucker2000hardware}
Daniel~F Zucker, Ruby~B Lee, and Michael~J Flynn.
\newblock Hardware and software cache prefetching techniques for mpeg
  benchmarks.
\newblock {\em IEEE Transactions on Circuits and Systems for Video Technology},
  10(5):782--796, 2000.

\end{thebibliography}

\appendix
\section{Appendix}
\label{sec:Appendix}
\subsection{LSTM Equations}
\label{sec:LSTM Equations}
We list LSTM equations below for completeness:
\begin{equation}
\label{eq: LSTM_input_gate}
i_t = \sigma(w_i[h_{t-1}, x_t] + b_i),
\end{equation}
\begin{equation}
\label{eq: LSTM_forget_gate}
f_t = \sigma(w_f[h_{t-1}, x_t] + b_f),
\end{equation}
\begin{equation}
\label{eq: LSTM_output_gate}
o_t = \sigma(w_o[h_{t-1}, x_t] + b_o),
\end{equation}

\begin{equation}
\label{eq: LSTM_cell_state_1}
\hat{c}_t = tanh(w_c[h_{t-1}, x_t] + b_c),
\end{equation}
\begin{equation}
\label{eq: LSTM_cell_state_2}
c_t = f_t * c_{t-1} + i_t * \hat{c}_t,
\end{equation}
\begin{equation}
\label{eq: LSTM_cell_state_3}
h_t = o_t * tanh(c_t),
\end{equation}

Where $x_t$ is the input, $h_{t-1}$ is the hidden state at time-step $t-1$, $i_t$ is the input gate, $f_t$ is the forget gate, and $o_t$ is the output gate. The weights for the input gate $i_t$, forget gate $f_t$, and the output gate $o_t$ are noted as $w_i, w_f, w_o$, respectively. Similarly, the biases for these gates are noted as $b_i, b_f, b_o$, respectively. Where $c_t$ is the cell state at time-step $t$, $\hat{c}_t$ is the candidate cell state at time-step $t$, and $\sigma$ is the sigmoid activation function. \\


\end{document}